
\newcount\eqnumber
\eqnumber=1
\def\chaphead{}

\def\new{\hbox{(\chaphead\the\eqnumber}\global\advance\eqnumber by 1}
\def\ref#1{\advance\eqnumber by -#1 (\chaphead\the\eqnumber
     \advance\eqnumber by #1 }
\def\first{\hbox{(\chaphead\the\eqnumber{a}}\global\advance\eqnumber by 1}
\def\last{\advance\eqnumber by -1 \hbox{(\chaphead\the\eqnumber}\advance
     \eqnumber by 1}
\def\eq#1{\advance\eqnumber by -#1 equation (\chaphead\the\eqnumber
     \advance\eqnumber by #1}
\def\eqnam#1{\xdef#1{\chaphead\the\eqnumber}}


\def\eqt#1{Eq.~({{#1}})}

\def\tcar{\futurelet\next\testnextcar}
\def\testnextcar{\ifhmode\ifcat\next.\else\ \fi\fi}

\def\Dt{\spose{\raise 1.5ex\hbox{\hskip4pt$\mathchar"201$}}} 

\def\a37{\hbox{$^{37}$Ar}}
\def\ar37{\hbox{$^{37}$Ar}}
\def\argon40{\hbox{$^{40}$Ar}}
\def\as71{\hbox{$^{71}$As}}
\def\b8{\hbox{$^{8}$B}}
\def\be7{\hbox{$^{7}$Be}}
\def\beril8{\hbox{$^8$Be}}
\def\boron11{\hbox{$^{11}$B}}

\def\br81{\hbox{$^{81}$Br}}
\def\carb12{\hbox{$^{12}$C}}
\def\ca37{\hbox{$^{37}$Ca}}
\def\car11{\hbox{$^{11}$C}}
\def\carbon13{\hbox{$^{13}C$}}

\def\chr51{\hbox{$^{51}$Cr}}
\def\cl37{\hbox{$^{37}$Cl}}
\def\cm2{\hbox{cm$^{2}$}}

\def\cu65{\hbox{$^{65}$Cu}}
\def\day-1{\hbox{day$^{-1}$}\tcar}

\def\dt{\spose{\raise 1.0ex\hbox{\hskip2pt$\mathchar"201$}}} 
\def\f17{\hbox{$^{17}$F}}

\def\ga71{\hbox{$^{71}$Ga}}
\def\gacl3{\hbox{GaCl$_3$}}
\def\ge71{\hbox{$^{71}$Ge}}
\def\gecl4{\hbox{GeCl$_4$}}
\def\geh4{\hbox{GeH$_4$}}

\def\h2{\hbox{$^{2}$H}}
\def\he3{\hbox{$^{3}$He}}

\def\helium4{\hbox{$^{4}$He}}

\def\hy1{\hbox{$^{1}$H}}
\def\ind115{\hbox{$^{115}$In}}
\def\iodine127{\hbox{$^{127}$I}}
\def\k37{\hbox{$~^{37}$K}}
\def\k40{\hbox{$^{40}$K}}
\def\kay40{\hbox{$~^{40}$K}}

\def\kr81{\hbox{$^{81}$Kr}}

\def\li7{\hbox{$^{7}$Li}}
\def\lithium8{\hbox{$^{8}$Li}}

\def\mo98{\hbox{$^{98}$Mo}}

\def\n13{\hbox{$^{13}$N}}
\def\nitrogen14{\hbox{$^{14}$N}}

\def\o15{\hbox{$^{15}$O}}
\def\ok2{\hbox{$~^{2}$K}}
\def\ox16{\hbox{$^{16}$O}}
\def\oxygen16{\hbox{$^{16}$O}}

\def\phib8{\hbox{$\phi(\b8)$}}
\def\phibe7{\hbox{$\phi(\be7)$}}

\def\phin13{\hbox{$\phi(\n13)$}}
\def\phio15{\hbox{$\phi(\o15)$}}

\def\sec-1{\hbox{sec$^{-1}$}}
\def\sec{\hbox{\rm sec}\tcar}

\def\sn115{\hbox{$^{115}$Sn}}

\def\tc98{\hbox{$^{98}$Tc}}
\def\tl205{\hbox{$^{205}$Tl}}

\def\va51{\hbox{$^{51}$V}}

\def\xenon127{\hbox{$^{127}$Xe}}

\def\zn65{\hbox{$^{65}$Zn}}

\def\ea37{\a37}
\def\ear37{\ar37}
\def\eargon40{\argon40}
\def\eb8{\b8}
\def\ebe7{\be7}
\def\eberil8{\beril8}
\def\eboron11{\boron11}
\def\ebr81{\br81}
\def\ec12{\carb12}
\def\eca37{\ca37}
\def\ecar11{\car11}
\def\ecarbon13{\carbon13}

\def\echr51{\chr51}
\def\ecl37{\cl37}

\def\ecu65{\cu65}

\def\eeminus{\eminus}
\def\eeminus\rm {e^-}

\def\ega71{\ga71}
\def\ege71{\ge71}

\def\eh2{\h2}
\def\ehe3{\he3}

\def\ehelium4{\helium4}

\def\ehy1{\hy1}
\def\ein115{\ind115}
\def\ek40{\k40}
\def\ekay40{\kay40}
\def\ekr81{\kr81}
\def\eli7{\li7}
\def\elithium8{\lithium8}

\def\eminus{\hbox{$\rm e^-$}}
\def\emo98{\mo98}
\def\emo98{\mo98}

\def\en13{\n13}
\def\enitrogen14{\nitrogen14}

\def\eo15{\o15}
\def\eoxygen16{\oxygen16}

\def\ephib8{\phi(\b8)}
\def\ephibe7{\phibe7}

\def\ephin13{\phin13}
\def\ephio15{\phio15}

\def\esn115{\sn115}

\def\etc98{\tc98}

\def\eva51{\va51}

\def\ezn65{\zn65}

\def\gtorder{\mathrel{\raise.3ex\hbox{$>$}\mkern-14mu
             \lower0.6ex\hbox{$\sim$}}}
\def\ltorder{\mathrel{\raise.3ex\hbox{$<$}\mkern-14mu
             \lower0.6ex\hbox{$\sim$}}}
\tolerance=10000
\baselineskip=1.5\normalbaselineskip
\centerline{\bf Recent Work on Standard Solar Models}
\bigskip\medskip
\centerline{John N. Bahcall}
\medskip
\centerline{Institute for Advanced Study, Princeton, New Jersey, 08540}
\bigskip\medskip\smallskip

{\bigskip\parindent=40pt
\centerline{\bf ABSTRACT}\medskip\narrower
Recent results on standard solar models are reviewed.
I shall summarize briefly three of the
themes that I stressed at the Neutrino '92 Conference: 1) Different solar model
codes give the same answers when the same input data are used; 2)
Improved calculations of standard solar models include helium diffusion,
the Livermore radiative opacity, the meteoritic iron abundance, and a
variety of other corrections; and 3) There are a few basic rules that
should be followed in using standard solar models.
At the Neutrino '92 Conference,
I reviewed in more detail the recent work on standard solar models by
Marc Pinsonneault and myself.  This work has by now appeared in print
(Rev. Mod. Phys. 64, 885, 1992, hereafter Paper I,
and ApJ Letters, 69, 717, 1992, Paper II).
Therefore, there is no need for me to repeat the details here.
\bigskip}

\noindent
1. DIFFERENT CODES GIVE THE SAME RESULTS FOR THE SAME INPUT DATA
\vskip.5truecm
Since the first accurate solar
model was derived in 1962, a number of different codes have been
developed to
calculate precise solar neutrino fluxes It was pointed out on page 78 of {\it
Neutrino Astrophysics}, Cambridge University Press (1989)
that the results of nine different codes
developed between 1962 and 1989 were consistent to within about 10\%
for the ${\rm ^8B}$ neutrino flux
when the same input parameters were used or when corrections were made
for the differences in input parameters.
References to the various calculations are given on the just-cited page 78.

Some authors have not been aware of this previously-established
convergence of calculated fluxes and
attributed differences between fluxes calculated using
some contemporary solar models to
different stellar physics.  Historically, a few examples of program
errors have been discovered, but these have all been
relatively minor in
their effects.  In fact, the contemporary
differences are essentially all
due to
different choices in input data.  This result was demonstrated in a
step-by-step comparison of recently-derived models in Section~4 of Paper I.

The final result of this detailed comparison is that, with the same
input parameters,
the standard solar
models of Bahcall and  Ulrich; of Sienkiewicz, Bahcall, and  Paczy\'nski;
of Turck-Chi\`eze, Cahen, Cass\'e, and  Doom; of
Sackmann, Boothroyd, and Fowler; and the current Yale model
(Bahcall-Pinsonneault)
all predict event rates for the chlorine experiment that are
the same within $0.2$ SNU.
The Sackmann, Boothroyd, and Fowler results are also apparently in
good agreement with the other calculations but Boothroyd has
informed me of
approximations that were embodied in their stellar evolution code
that
prevent a precise quantitative comparison.
\vskip1truecm
\noindent
2. IMPROVEMENTS IN SOLAR MODELS
\vskip.5truecm
A number of improvements have been made in the input data for standard
solar models over the past four years.
I will first discuss the effects of helium diffusion, radiative opacity, and
iron abundance, all of which have comparable effects (about 1 SNU) on
the predicted event rate in the chlorine experiment.

The new physics that is most difficult to include in stellar evolution
codes is the diffusion of
helium with respect to hydrogen in the solar interior.
Convenient formulae were derived for this effect by Bahcall and Loeb,
but these formulae still cause complications in
the stellar evolution codes because diffusion
mixes spatial and time derivatives. So far, the only precise solar
models including this effect were presented in Paper I.
Helium diffusion
increases the predicted event rate in the chlorine experiment by about
0.8 SNU.
That is to say, if you want to compare two otherwise similar solar
models in their predicted event rates, you must add SNUs to the predictions
of the model that does not include helium diffusion:\ 0.8 SNU for the chlorine
experiment and 5 SNU for the gallium experiment.

A major improvement in the radiative opacities was achieved through the
calculations of Iglesias and Rogers, whose results are
incorporated in the now widely used Livermore Opacities.  These
opacities give better agreement in comparing stellar model predictions
with observations of $\beta$ Cephei pulsations, RR Lyrae masses, double
mode Cepheids, double mode $\delta$ Scutti stars, Li abundance in the
Hyades cluster, the depth of calculated convective cores, and the
luminosity-effective temperature relation for O and B stars.
The use of Livermore rather than Los Alamos opacities increases the
calculated event rate by about 1 SNU for the chlorine experiment.

There has long been a major uncertainty in the predicted event rates due
to the different values inferred observationally for the abundance of
iron on the solar surface.  In Paper I, we adopted--following recent
reanalyses of the reliability of the photospheric  abundances--the lower iron
abundance obtained by studying meteorites
rather than the higher abundance obtained by analyzing the solar
photosphere.  This choice reduces the event rate by 1.3 SNU.

Paper I also describes a large number of less important corrections
including more accurate nuclear cross sections for several fusion
reactions, a more detailed electron screening calculation (although a
small correction may still need to be made in this area, see the
discussion by Turck-Chi\`eze in this volume), and a few programming
improvements.

The predicted rate for the chlorine experiment given in Paper I is
\eqnam{\chlorinetheory}
$$
<\phi \sigma >_{\rm Cl,~ theory} ~=~ (8.0 \pm 1.0) ~ {\rm SNU},
\eqno\new)
$$
where I have reluctantly quoted here--in deference to my experimental
friends--an approximate $1\sigma$ error (instead of my preferred
``total theoretical error'', which includes $3 \sigma$ errors for all
measured quantities).
The difference between the experimental rate of 2 SNU presented
at this conference and the theoretical rate given in
\eqt{\chlorinetheory} constituted for two decades the ``solar neutrino
problem.''

The predicted rate for the gallium experiment is
\eqnam{\GaTheory}
$$
\sum(\phi\sigma)_{\rm Ga} ~=~ 131.5_{- 6}^{+ 7}\  {\rm SNU},
\eqno\new)
$$
again approximate $1 \sigma$ errors.
The GALLEX result is about $2 \sigma$  less than the calculated rate;
the SAGE result is more than $2 \sigma$ less than the theoretical value.

The calculated flux of $^8$B neutrinos is

\eqnam{\Yes8Bflux}
$$
\phi (^8{\rm B})~=~5.7 (1 \pm 0.15) \times 10^6~{\rm cm}^{-2}{\rm
s}^{-1}~,
\eqno\new)
$$
which is approximately twice the measured Kamiokande rate reported at
this conference.
\vskip1truecm
\noindent
3. HOW TO USE STANDARD SOLAR MODELS
\vskip.5truecm
Physicists are used to comparing the results of different experiments
that measure the same or related quantities.  We are not used to
comparing different solar models.  However, for the next few years
we must continue to use the predictions
of accurate solar models to interpret what we have learned from
solar neutrino experiments.
With the next generation of experiments (SNO, Super-Kamiokande, ICARUS),
precise measurements of the $^8$B energy spectrum and
the neutral current to charged current ratio will allow us to
decide if there is new physics required independent of solar model
calculations.
But, these results will presumably not become available until sometime
in 1996 or later.

What should one do to be an informed consumer of solar models?  Here is a
list of questions that you should ask about solar models.

\noindent{\bf (1.) Does the model yield the same results
(to about $\pm 0.1$ SNU
for the chlorine experiment) as other standard models  with the same
input data?}

Paper I verifies that this condition is satisfied for all
of the contemporary models discussed at this conference.

\noindent{\bf (2.) Have the authors used the best published values (and the
quoted experimental errors) on all measured quantities?}

Most people
calculating stellar models do not bother to continually upgrade their
calculations as more accurate measurements of, e.g., nuclear quantities
become available.  Some models discussed at this conference were not
adjusted for the recent thorough reanalysis by the Cal Tech group of
the ${\rm ^8B}$ production cross section (with a derived error).
An exportable nuclear energy generation
and neutrino production subroutine that makes this easy to accomplish is
can be obtained from the author.

\noindent{\bf (3.) Does the model include the best-available
theoretical input data?}

 Of particular importance are
radiative opacities, element diffusion, and equation of state.
Of the models discussed in this conference,
only the models in Paper I include the improvements due to
helium diffusion and to the Livermore radiative opacities. Most of the
models include in the adopted equation of state the dominant correction
caused by the Debye-Huckel effect.

\noindent{\bf (4.)  Are the errors well defined and equal to the published
errors determined by the experimentalists?}

For anyone interested in
the subject of errors on solar models, Section 8 of Paper I contains
a current discussion of the best-estimates of the errors and a
comparison of how different errors have been estimated.
Some authors have quoted subjective uncertainties that are
larger than the values given here and in Paper I.
I believe that
the well-defined errors derived
in Paper I and in this report are conservatively
large.
The reader is invited to consider Section 8 of Paper I and to
judge on the basis of the detailed discussion
whether the quoted uncertainties are
indeed conservative.
\vskip1truecm
\noindent
4. CLOSING COMMENTS
\vskip.5truecm
We need more solar neutrino experiments.  It is possible that we as a
community may have been lucky and have stumbled onto some important
microscopic physics
while trying to test accurately stellar evolution theory.
But, the available results depend upon a small number of experiments
with redundancy only in the gallium experiments and with a continued
reliance on the standard solar model calculations.
Unless the gallium
results converge together on a value that does not exceed 80 SNU
(cf.~chapter 11 of {\it Neutrino Astrophysics}) even with the
experimental errors, we will probably have to wait for the next
generation of experiments (SNO, Super-Kamiokande, Borexino, and ICARUS)
before we can prove that we have learned new physics or new
astrophysics.
\vskip1truecm
\noindent
{\bf ACKNOWLEDGMENTS}
\vskip.5truecm
This work was supported in part by the NSF via grant
PHY-92-45317 at I.A.S..

\bye